\documentclass[aps,prl,reprint,groupedaddress,showpacs,amsmath]{revtex4-1}

\usepackage{graphics}
\usepackage{dcolumn}

\begin{document}

\title{Quantum Langevin model for exoergic ion-molecule reactions and inelastic processes}


\author{Bo Gao}
\email[]{bo.gao@utoledo.edu}
\affiliation{Department of Physics and Astronomy,
	Mailstop 111, University of Toledo, 
	Toledo, Ohio 43606,
	USA}

\date{January 31, 2011}

\begin{abstract}

We presents a fully quantal version of the Langevin
model for the total rate of exoergic ion-molecule reactions
or inelastic processes. The model, which
is derived from a rigorous multichannel quantum-defect formulation
of bimolecular processes, agrees with the classical Langevin model at
sufficiently high temperatures. It also gives the first analytic
description of ion-molecule reactions and inelastic processes
in the ultracold regime where the quantum nature of the relative
motion between the reactants becomes important.

\end{abstract}


\maketitle

The classical Langevin model for exoergic ion-molecule 
reactions \cite{lan05,fer06} is one of the most
fundamental and powerful results in the theory of reactions.
It has been shown to be applicable to a variety of systems 
and over a wide range of temperatures \cite{Gioumousis1958,fer06}.
As a model based on the long-range interaction,
it can be expected to be more accurate the lower the
temperature \cite{geo05}, until one reaches a regime where quantum effects,
more specifically the quantum effects associated with the relative
motion of the reactants, become important.

An experimental study of reactions in this temperature regime,
often referred as the ultracold regime, has recently been realized
for neutral-neutral reactions in a landmark experiment by the
JILA group \cite{osp10}.
While it is not yet realized for charge-neutral
systems, the growing ability of making
cold molecular samples \cite{ni08,*dan10}
and manipulating cold ions \cite{gri09,*zip10,*zip10b}, implies that
it may soon become a reality. 
A sample process, likely among the first investigated, 
would include the type
\begin{subequations}
\label{eq:imrex}
\begin{eqnarray}
A^++B_2 &\rightarrow& AB+B^+ \;,\\
	&\rightarrow& A+B_2^+ \;,\\
	&\rightarrow& (AB)^++B \;,
\end{eqnarray}	
\end{subequations}
where all reactions can be expected to be exoergic if the ionization
potential of atom $A$ is considerably greater than that of atom $B$.
They can proceed with substantial rates even in the limit of
zero temperature as ion-molecule interactions are generally 
expected to be barrierless at the short range
\footnote{The applicability of the Langevin model only
requires that one of the processes to be exoergic and barrierless.}. 

The theory presented here, to be called the quantum Langevin (QL) model 
for the total rate of exoergic ion-molecule 
reactions or inelastic processes, gives the first 
theoretical prediction on where and how the quantum effects 
come into play and how the resulting behavior
deviates from the classical Langevin model. It is another 
application of a new quantum framework for reactions \cite{gao10b}
that differs considerably from existing formulations 
\cite{[][{ and references therein.}]Hu2006}, and is used
here to further illustrate the concepts behind the theory.

In a conventional quantum theory of reactions \cite{Hu2006}, 
little can be known about a reaction without a detailed
knowledge of the potential energy surface (PES), the accuracy of
which is often insufficient for quantitative predications, especially
at low temperatures. This has been true even for ultracold atom-atom and 
ion-atom interactions, where, to the best of our knowledge, no \textit{ab initio}
PES for alkali metal systems has ever been sufficiently accurate to predict 
the scattering length. All potentials had to be modified by incorporating a substantial
amount of spectroscopic data (see, e.g., Ref.~\cite{Ferber2009}). 
The same issue becomes \textit{much} more severe for multidimensional
PES in reactions, and will remain so for many years to come.
This difficulty, coupled with the exponential growth of the
Hilbert space beyond two-body \cite{Quemener2005}, has limited the conventional
approach to a few simple systems such as D+H$_2$, with little
hope for more complex systems.

The multichannel quantum-defect theory for reactions 
(MQDTR), as outline in Ref.~\cite{gao10b}, with important 
motivations and ingredients that come
before it \cite{osp10,jul09,idz10,que10,idz10b}, 
offers a different perspective on reactions and
inelastic processes.
It comes from a much wider assertion that much can be known
about a quantum system, specifically its behavior around a fragmentation
threshold, simply from the \textit{types} of long-range
interactions among its constituents. Whatever not yet known can
be characterized by a few energy-insensitive parameters, which can be 
further determined from a few experimental measurements without 
any knowledge of the short-range interaction, or even the
strength of the long-range interaction. This physical picture,
which goes back to the original quantum-defect theory for 
the Coulomb interaction \cite{gre79,*sea83},
has been well established
in recent years for atom-atom \cite{[][{ and references therein.}]gao05a,gao08a}
and ion-atom interactions \cite{gao10a}, and to a lesser degree 
for few-atom \cite{kha06} and many-atom systems \cite{gao04a,*gao05b}.
Its realization for reactions \cite{gao10b}, as is further illustrated
in this work, frees the theory from being held hostage by the details 
of PES, while ready to take advantage of them when they are
available. It also resolves a conceptual disparity in existing theories
of reactions. While many classical models \cite{fer06,geo05} 
are based on the recognition of the
importance of the long-range interaction, the same physical concept
gets lost in the conventional quantum formulations \cite{Hu2006}.
It is primarily due to this omission that they have missed
the universality and the simplicity in reactions that have
been uncovered in the landmark JILA experiment \cite{osp10}.

The QL model for ion-molecule reactions or inelastic processes,
to be presented here, is a 
special case of a more general QL model \cite{gao10b} that is 
formally applicable to bimolecular processes with arbitrary $-C_n/R^{n}$ ($n>2$) 
type of long-range interaction in the
entrance channel. Other than the exponent $n$, the most important
characteristic of such a potential is its length scale
$\beta_n=(2\mu C_n/\hbar^2)^{1/(n-2)}$, where $\mu$ is the reduced
mass in the entrance channel. It determines the scale parameters
for other relevant physical observables, such as the energy, with a scale
of $s_E=(\hbar^2/2\mu)(1/\beta_n)^2$, the temperature, with a
scale of $s_E/k_B$, and the rate of reactions, with a scale of
$s_K=\pi\hbar\beta_n/\mu$ \cite{gao10b}.
For ion-molecule interactions, $n=4$ and $C_4=\alpha q^2/2$,
corresponding to the polarization potential, with
$\alpha$ being the average polarizability of the molecule
and $q$ the charge of the ion. The realization of the QL model 
for this class of systems is made possible
by a recent reformulation of the quantum-defect theory (QDT)
for $-1/R^4$ potential \cite{gao10a}, which gives, in particular,
analytic results for the quantum transmission probability to be used
in this work.

In Ref.~\cite{gao10b}, we have shown that under the Langevin
assumption, corresponding to the assumption of no reflection by
the inner potential, the total rate of reactions and inelastic
processes follows a universal behavior uniquely determined by
the exponent $n$ characterizing the type of long-range interaction
in the entrance channel. Different systems with the same type
of potentials differ from each other only in scaling.
Specifically, the rate constant for the total
rate of reactions and inelastic processes can be written as
\begin{equation}
K(T) = s_K {\mathcal K}^{(n)}(T_s) \;,
\end{equation}
where $s_K$ is the rate scale defined earlier, and
${\mathcal K}^{(n)}(T_s)$ is a universal function of the
scaled temperature, $T_s = T/(s_E/k_B)$, given by
\begin{equation}
{\mathcal K}^{(n)}(T_s) = \frac{2}{\sqrt{\pi}}
	\int_0^\infty dx\: x^{1/2} e^{-x}
	{\mathcal W}^{(n)}(T_s x)\;.
\label{eq:urf}
\end{equation}
Here ${\mathcal W}^{(n)}(\epsilon_s)$ is a scaled total rate
before thermal averaging. It depends on energy only through the
scaled energy $\epsilon_s = \epsilon/s_E$, and has contributions
from all partial waves
\begin{equation}
{\mathcal W}^{(n)}(\epsilon_s) = \sum_{l=0}^\infty {\mathcal W}^{(n)}_{l}(\epsilon_s) \;,
\end{equation}
where ${\mathcal W}^{(n)}_{l}$ is a scaled partial rate
given by
\begin{equation}
{\mathcal W}^{(n)}_{l}(\epsilon_s)
	=(2l+1){\mathcal T}^{c(n)}_l(\epsilon_s)/\epsilon_s^{1/2} \;,
\label{eq:prf}	
\end{equation}
in which ${\mathcal T}^{c(n)}_l(\epsilon_s)=|t^{(oi)}_l(\epsilon_s)|^2$
is the quantum transmission probability through the long-range
potential at the scaled energy $\epsilon_s$ and 
for partial wave $l$ \cite{gao08a}.

For $n=4$, corresponding to the ion-molecule interaction, 
the quantum transmission probability,
${\mathcal T}^{c(4)}_l(\epsilon_s)$, which
is the only quantity required to determine the universal
rate function in the QL model, can be found analytically
as a part of the QDT for $-1/R^4$ 
potential \cite{gao08a,gao10a}.
The result is
\begin{equation}
{\mathcal T}^{c(4)}_l(\epsilon_s) = \frac{2M_{\epsilon_s l}^2[1-\cos(2\pi\nu)]}
	{1-2M_{\epsilon_s l}^2\cos(2\pi\nu)+M_{\epsilon_s l}^4}\;,
\label{eq:tp4}	
\end{equation}
where $\nu$ is the characteristic exponent 
and $M_{\epsilon_s l}$ is one of the universal QDT functions 
for the $-1/R^4$ potential, both of which are as given 
in Ref.~\cite{gao10a}.

\begin{figure}
\scalebox{0.4}{\includegraphics{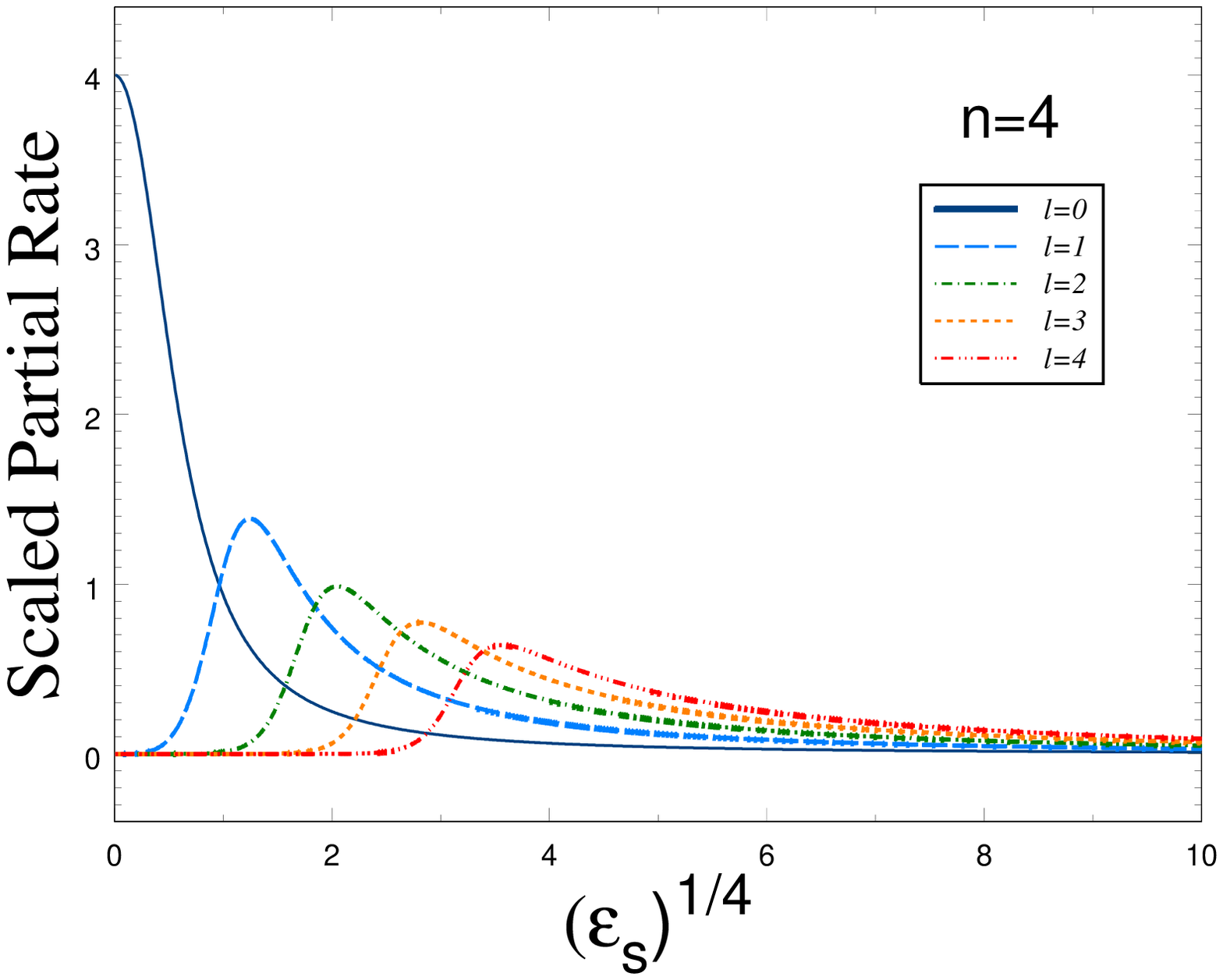}}
\caption{(color online) The scaled partial rates
${\mathcal W}^{(n)}_{l}(\epsilon_s)$ for $n=4$, corresponding to
$-1/R^4$ type of interaction in the entrance channel.
\label{fig:WlQL4}}
\end{figure}
\begin{figure}
\scalebox{0.4}{\includegraphics{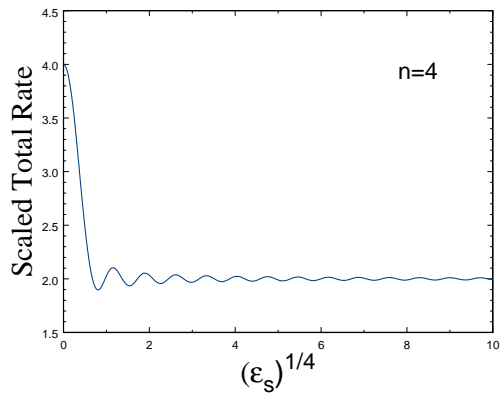}}
\caption{(color online) The scaled total rate
${\mathcal W}^{(n)}(\epsilon_s)$ for $n=4$, corresponding to
$-1/R^4$ type of interaction in the entrance channel.
\label{fig:WQL4}}
\end{figure}
Figure~\ref{fig:WlQL4} illustrates the resulting scaled partial rates
${\mathcal W}^{(n)}_{l}$ for $n=4$. Figure~\ref{fig:WQL4} 
shows the corresponding total rate ${\mathcal W}^{(n)}$.
We note that the oscillatory structure in the total rate is neither
a resonance nor an interference phenomenon. It is instead a result of the
quantization of angular momentum, 
with contributions from a discrete set of
partial waves peaking at different energies, as illustrated in 
Fig.~\ref{fig:WlQL4}.
The ${\mathcal W}^{(n)}$ is related to the 
total reactive and inelastic cross section, $\sigma_{\mathrm{ur}}$, by
$\sigma_{\mathrm{ur}} = (\pi\beta_{n}^2)\Sigma^{(n)}_{\mathrm{ur}}(\epsilon_s)$,
in which
$
\Sigma^{(n)}_{\mathrm{ur}}(\epsilon_s) = {\mathcal W}^{(n)}/\epsilon_s^{1/2} \;
$
is the scaled total inelastic and reactive cross section.
We note that the oscillatory structure in ${\mathcal W}^{(n)}$
is much less visible in the cross section, making it difficult
to detect experimentally.

\begin{figure}
\scalebox{0.4}{\includegraphics{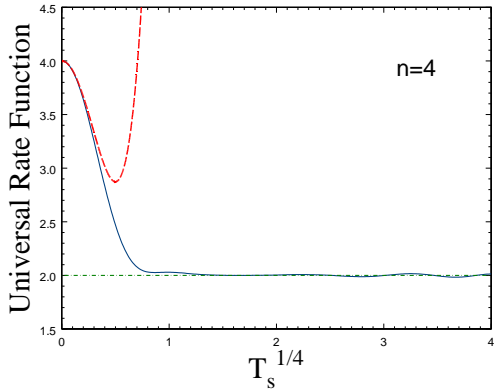}}
\caption{(color online) The universal rate function
${\mathcal K}^{(n)}(T_s)$ for $-1/R^4$ type of interaction 
in the entrance channel (solid line).
The dash-dot line represents the prediction of the classical
Langevin model, as given by Eq.~(\ref{eq:Langevin}).
The dashed line represents the QDT expansion as given by Eq.~(\ref{eq:K4exp}).
\label{fig:urf4}}
\end{figure}
Figure~\ref{fig:urf4} illustrates the universal rate function
${\mathcal K}^{(n)}(T_s)$ for $n=4$,
which is the thermal average of the total rate shown in Fig.~\ref{fig:WQL4}. 
In the ultracold regime of $T_s\ll 1$, simpler analytic
formulas for rates and cross sections can be derived
using the QDT expansion \cite{gao09a} of the transmission probability.
For small scaled energies $\epsilon_s$, 
it can be shown from the QDT for $n=4$ \cite{gao10a} that
\begin{equation}
{\mathcal T}^{c(4)}_{l=0} = \frac{4\bar{a}^{(4)}_{sl=0}\epsilon_s^{1/2}}
	{(1+\bar{a}^{(4)}_{sl=0}\epsilon_s^{1/2})^2}+O(\epsilon_s^{5/2}) \;,
\label{eq:tex1}
\end{equation}	
\begin{equation}
{\mathcal T}^{c(4)}_{l=1} = \frac{4\bar{a}^{(4)}_{sl=1}\epsilon_s^{3/2}}
	{(1+\bar{a}^{(4)}_{sl=1}\epsilon_s^{3/2})^2}+O(\epsilon_s^{7/2}) \;,
\end{equation}
and
\begin{equation}
{\mathcal T}^{c(4)}_{l\ge 2} = 4\bar{a}^{(4)}_{sl}\epsilon_s^{l+1/2}
	+O(\epsilon_s^{l+5/2}) \;,
\label{eq:tex3}
\end{equation}
where 
\begin{equation}
\bar{a}^{(4)}_{sl} = \frac{(2l+1)^2}{[(2l+1)!!]^4} \;,
\end{equation}
is called the scaled mean scattering lengths for a $-1/R^4$ potential, 
after similar quantities for $-1/R^6$ potential \cite{gao09a}. 
We note that such analytic expansions
could not have been derived from either the WKB theory \cite{mer98}, 
or the top-of-barrier analysis \cite{war00}, as to be discussed
in more details elsewhere \cite{li10}.
Substituting Eqs.~(\ref{eq:tex1})-(\ref{eq:tex3}) into
Eqs.~(\ref{eq:urf})-(\ref{eq:prf}) gives the expansion of 
${\mathcal K}^{(4)}(T_s)$ in the ultracold regime of $T_s\ll 1$,
\begin{eqnarray}
{\mathcal K}^{(4)}(T_s) &=& 4\bar{a}^{(4)}_{sl=0}
	-\frac{16(\bar{a}^{(4)}_{sl=0})^2}{\sqrt{\pi}}T_s^{1/2} \nonumber\\
	& &+18\left[(\bar{a}^{(4)}_{sl=0})^3+\bar{a}^{(4)}_{sl=1}\right]T_s 
	+ O(T_s^{2})\;.
\label{eq:K4exp}	
\end{eqnarray}
Here $\bar{a}^{(4)}_{sl=0}=1$ and $\bar{a}^{(4)}_{sl=1}=1/225$
are the scaled mean scattering lengths for $l=0$ and $l=1$,
respectively. A comparison of this QDT expansion with the
exact result is shown in Fig.~\ref{fig:urf4}.

At high temperatures as characterized by $T_s\gg 1$, 
it is straightforward to show, from the semiclassical limit
of the transmission probabilities \cite{gao08a}, that
\begin{equation}
{\mathcal K}^{(4)}(T_s)\sim 2 \;,
\label{eq:Langevin}
\end{equation}
in agreement with the classical Langevin model \cite{lan05,fer06}.
Figure~\ref{fig:urf4} shows that the transition from quantum
to semiclassical behavior occurs over a temperature range of
$s_E/k_B$, which we generally refers as the van der Waals 
temperature scale. The rate goes from $4s_K$ at the threshold
to approximately the Langevin rate of $2s_K$ beyond $s_E/k_B$.
Figure~\ref{fig:urf4} also shows that the oscillatory structure present 
in ${\mathcal W}^{(n)}$ has mostly been washed out by thermal averaging.

\begin{table}
\caption{Sample scale parameters for ion-molecule 
reactions and inelastic processes.
The $\beta_4=(2\mu C_4/\hbar^2)^{1/2}$ is the length scale
associated with the polarization potential $-C_4/R^4$,
where $C_4=\alpha q^2/2$ with $\alpha$ being the average polarizability
of the molecule. 
$s_E/k_B=(\hbar^2/2\mu)(1/\beta_4)^2/k_B$ 
is the corresponding temperature scale. 
$s_K=\pi\hbar\beta_4/\mu$ is the rate scale corresponding to $\beta_4$.
It is given here in units of $10^{-9}\;\mathrm{cm}^3\mathrm{s}^{-1}$.
\label{tb:scales}}
\begin{ruledtabular}
\begin{tabular}{lrrrr}
System                       & $\alpha$ (a.u.)     & $\beta_4$ (a.u.) & $s_E/k_B$ (K)   & $s_K$ \\
\hline
D$^{+}$-$^{1}$H$_2$          & 5.41\footnotemark[1]& 99.7             & $8.65\times 10^{-3}$ & 1.05       \\
$^1$H$^{+}$-$^{7}$Li$_2$     & 216\footnotemark[2] & 608              & $2.49\times 10^{-4}$ & 6.83       \\
$^{7}$Li$^+$-$^{87}$Rb$_2$   & 553\footnotemark[2] & 2610             & $1.90\times 10^{-6}$ & 4.08      \\
$^{138}$Ba$^+$-$^{87}$Rb$_2$ & 553\footnotemark[2] & 8800             & $1.45\times 10^{-8}$ & 1.21      \\
$^{40}$Ca$^+$-$^{133}$Cs$_2$ & 675\footnotemark[2] & 6540             & $5.83\times 10^{-8}$ & 1.99      \\
$^{40}$Ca$^+$-$^{40}$K$^{87}$Rb & 526\footnotemark[2] & 5400             & $9.78\times 10^{-8}$ & 1.88      \\
\end{tabular}
\end{ruledtabular}
\footnotetext[1]{From Ref.~\cite{kol67}.}
\footnotetext[2]{From Ref.~\cite{Tarnovsky1993,*dei08}.}
\end{table}
All scaled results can be put on absolute scales using a single parameter,
the average polarizability of the molecule $\alpha$.
It determines both the rate scale $s_K$ and the temperature scale
$s_E/k_B$. Sample scale parameters are given
in Table~\ref{tb:scales}.
They are chosen to illustrate that the meaning of the ultracold
regime, if defined as the range of temperatures over which
the quantum effects are important, is very different for different systems.
It covers a much broader temperature range for lighter systems 
than for heavier ones.
They are also chosen to imply that we expect the QL model 
to be applicable not only to nonpolar molecules,
but also to small polar molecules
such as KRb \cite{osp10}, with the main difference being that 
its upper range of applicability will be more limited 
\footnote{This issue, which is related to the ion-molecule interaction
at the next longest length scale where it is generally anisotropic,
is under much further investigation.}.
It is interesting to note from the table that despite wide
variations of temperature scales for different systems,
the rate scales are of the same order of magnitude,
$\sim 10^{-9}\;\mathrm{cm}^3\mathrm{s}^{-1}$,
which is roughly 100 times greater than those 
for neutral-neutral reactions \cite{gao10b}.

In conclusion, we have presented a fully quantal version of
the Langevin model for exoergic charge-neutral reactions and 
inelastic processes. It is a universal model in which different 
systems differ only in scaling, further illustrating the 
concept that even in a purely quantum theory, there are important 
aspects of reactions that
can be understood without detailed knowledge of the PES at
the short range. Such aspects include not only the total
rate of reactions and inelastic processes, presented here, 
but also the elastic cross section and the total cross section,
to be presented elsewhere.
For a state-to-state partial cross section, it can be shown
from the underlying MQDTR \cite{gao10b} that while its absolute value
requires the short-range PES, its energy dependence
can still be parametrized using the same universal transmission
probabilities ${\mathcal T}^{c(n)}_l(\epsilon_s)$.
We point out that the QL model is applicable not 
only to molecules in the ground or
low-lying states, but also to molecules in vibrationally highly excited
states and to atoms in selective Rydberg states 
(ones with significant quantum defect).
In such applications, the theory connects with 
quantum few-body physics \cite{bra06,*[][{ and references therein.}]gre10},
and provides a description of their behavior outside of
the so-called universal regime, a region 
that has been difficult to treat using other methods
because of the large number of open channels.
In a mathematical abstraction with even broader implications, 
the QL models presented here
and earlier in Ref.~\cite{gao10b} represents one type
of universal behaviors that can emerge whenever the number
of open channels in a set of coupled channel (or close-coupling) 
equations becomes large. 
It is our belief that uncovering and taking advantage 
of such universal behaviors will be the key to a more 
systemic understanding of quantum systems
beyond two-body. 
 
\begin{acknowledgments}
This work was supported by the NSF under 
the Grant No. PHY-0758042.
\end{acknowledgments}

\bibliography{chem,ionchem,ieatom,bgao,twobody,atomAtom,fewbody}

\end{document}